\documentclass{PoS}

\title{Ultra-Peripheral Collisions with gold ions in STAR}

\ShortTitle{Ultra-Peripheral Collisions with gold ions in STAR} 

\author{\speaker{Spencer R. Klein, for the STAR Collaboration}\\
       Lawrence Berkeley National Laboratory, Berkeley CA 94720 USA\\
      Author  E-mail: \email{srklein@lbl.gov}}

\abstract{In 2010, the STAR Collaboration collected 38 million low-multiplicity  triggers
for ultra-peripheral collision studies. We present new results
involving photonuclear interactions in ultra-peripheral relativistic heavy ion
collisions (UPCs), based on an analysis of 384,000 photoproduced pion
pairs, from $\rho$, $\omega$ and direct $\pi^+\pi^-$ pair production.The $\omega$
component is clearly visible through its interference with the $\rho$
peak.  The relative amplitudes of the three components have been measured, 
along with the phase angle between the $\rho$ and $\omega$ components.

Using a two-year (2010 and 2011) dataset, we explore higher mass final states, and observe a $\pi^+\pi^-$ state with a mass of $1653 \pm 10$ MeV and a width of $164\pm 15$ MeV (statistical errors only).  The state is at least roughly compatible with the $\rho_3(1690)$. 

The $\rho^0$ squared momentum transfer ($t$) spectrum exhibits a spectrum with both coherent and incoherent
prodution.   The coherent component shows two characteristic diffraction minima, with positions that are sensitive to
the hadronic radius of the gold nucleus.}

\FullConference{XXIV International Workshop on Deep-Inelastic Scattering and Related Subjects\\
		11-15 April, 2016\\
		DESY Hamburg, Germany}

\begin{document}
\section{Introduction}

Ultra-peripheral collisions (UPCs) allow us to study photoproduction and two-photon physics at hadron colliders, by taking advantage of the photon fields carried by relativistic nuclei.  In these collisions, the nuclei exchange one or more photons, but pass each other at impact parameters $b$ large enough so that no hadronic interactions occur.   The photon flux scales as the atomic number, $Z^2$, providing high photon-nucleus luminosity \cite{Bertulani:2005ru,Baltz:2007kq}.   UPCs have been studied at RHIC for many years, with measurments by STAR of $\rho^0$ photoproduction at energies from 62.4 to 200 GeV per nucleon, along with studies of purely electromagnetic $e^+e^-$ production and photoproduction of $4\pi$ final states.  The PHENIX collaboration studied $J/\psi$ production, and the beam loss monitors were used for a measurement of bound-free pair production.  These previous RHIC results were recently reviewed \cite{Klein:2015qna}.  

Because of the very large photon fluxes, two nuclei can exchange multiple photons.  This greatly simplifies UPC triggering.  One can select events where there is mutual Coulomb excitation, plus photoproduction of a  pion pair, greatly simplifying the trigger setup.  This analysis will consider two particular cases, pion pairs accompanied by mutual Coulomb excitation, with single neutron emission from each nucleus (1n1n) and excitation to any number of neutrons (XnXn). 

STAR detects charged particles with pseudorapidity $|\eta|<1.5$ in a large Time Projection Chamber (TPC) in a 0.5 T solenoidal magnetic field.    The TPC is surrounded by a Time-Of-Flight (TOF) system and an electormagnetic calorimeter.  Two Zero Degree Calorimeters (ZDCs) $\pm 18 $m downstream from the interaction point detect neutrons from nuclear breakup, and two Beam-Beam Counters (BBCs) detect charged particles with $2 < |\eta|<5$.  
Two upgrades have dramatically improved STARs capabilities to study UPCs.  First, the data acquisition system was upgraded with 'DAQ1000,' which raised the event readout rate by a factor of 10.  Second, the old central trigger barrel system was replaced with the TOF system, which allowed for more granular triggering.  In concert with the higher luminosity, these improvements allowed for a 100-fold increase in UPC data samples.  This writeup presents two analyses: a study of $\rho^0$, $\omega$ and direct $\pi\pi$ production using STAR data acquired in 2010, and a study of high mass $\pi\pi$ pairs using a combined data sample from 2010 and 2011.

\section{$\rho^0$, $\omega$ and direct $\pi\pi$}

The $\rho^0$, $\omega$ and direct $\pi\pi$ analysis used 38 million triggers recorded in 2010, selected from 1074 $\pm$ 107 $\mu$b$^{-1}$ of integrated luminosity.  The trigger selected events with one to five neutrons in each ZDC, between two and six charged particles seen in the TOF system and no hits in the BBCs.  The analysis selected pairs of tracks which were consistent with originating from a single vertex within 50 cm of the center of the interaction region. Each track was required to have at least 14 out of 45 normally possible hits in the TPC, and  a hit in the TOF system; the latter selected only tracks with $|\eta|<1.0$.  A total of 384,000 pairs were used in the analysis.  

The main backgrounds for this analysis were low-multiplicity hadronic interactions, other UPC interactions, beam-gas interactions and cosmic-ray muons, accompanied by an in-time mutual Coulomb excitation.   Most of the hadronic interaction background  was removed by subtracting the like-sign pairs from the oppositely charged pairs.   The other UPC interactions - electromagnetic production of $e^+e^-$ and photoproduction of $\omega$ followed by $\omega\rightarrow\pi^+\pi^-\pi^0$, with the $\pi^0$ missed, contributed a few percent of the pairs.

The reconstructed events are corrected for acceptance and detector efficiency using Monte Carlo events from the STARlight generator \cite{Klein:1999qj,Klein:1999gv}.   STARlight events were sent through a detailed GEANT model of the STAR detector, embedded in `zero bias' events to account for detector noise and event pileup, and then reconstructed with the same programs used on the actual data.  

\begin{figure}[t]
\includegraphics[width=0.8\columnwidth]{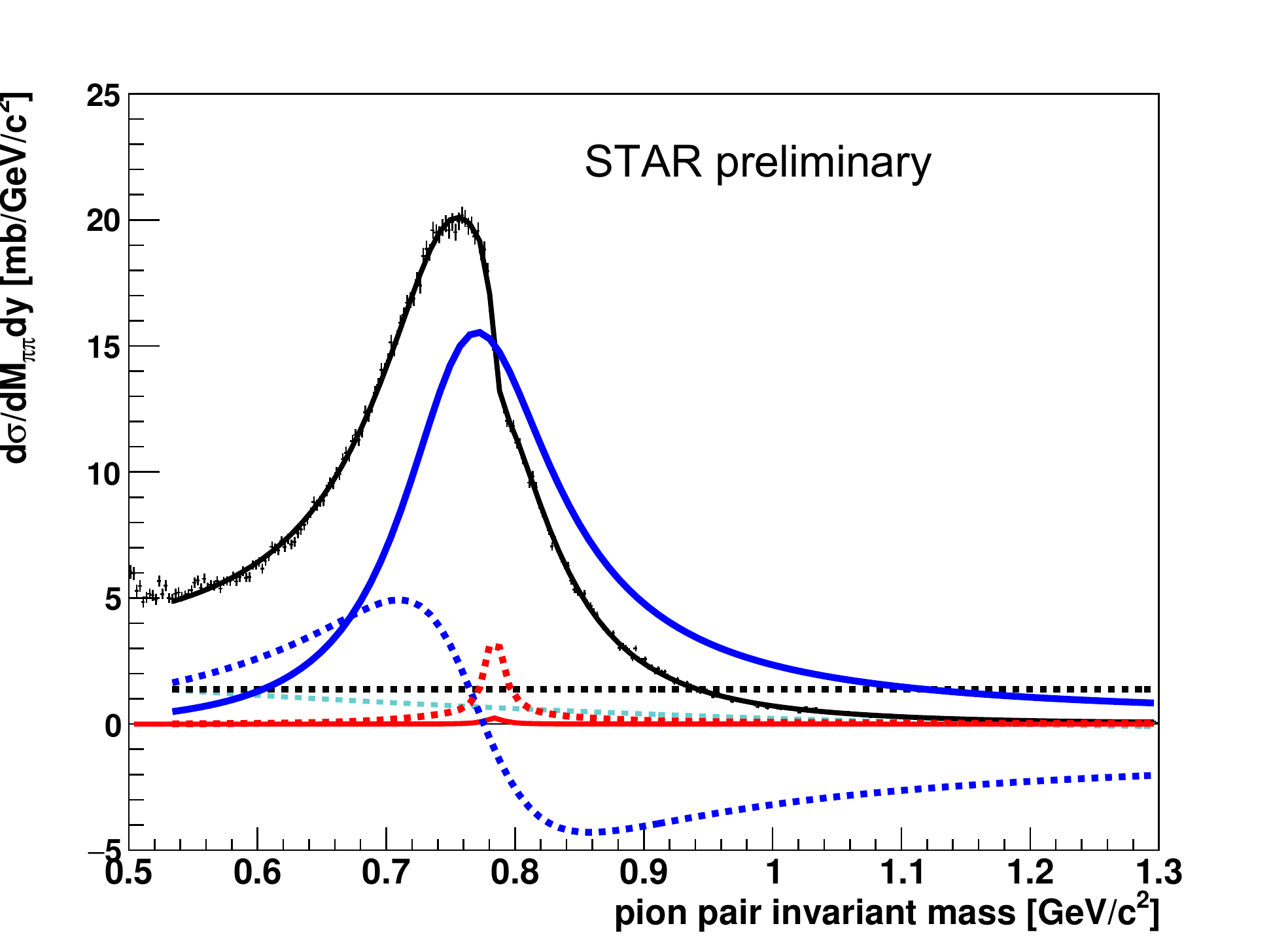} 
\label{fig:invMass}
\caption{STAR preliminary $\pi^+\pi^-$ invariant-mass distribution for events with $p_T < 100$ MeV/c, with 2.5 MeV/c$^2$ wide bins.  The black curve shows Eq. 2.1 
fit to the data. 
The $\rho$ Breit-Wigner is shown by the blue curve while the non-resonant pion pair component is shown with the black dashed curve.  The interference between non-resonant pion pairs and the $\rho$ meson is shown with a blue-dashed curve. The    $\omega$ component is shown in red, and the interference between $\rho$ and $\omega$ is shown with a red-dashed curve. The second order polynomial shown with a cyan-dashed curve is used to account for a remnant background.}
\end{figure}

Figure 1 shows the efficiency-corrected background-subtracted mass spectrum for the selected $\pi^+\pi^-$ pairs with $p_T < $100 MeV/c; the latter requirement selects coherent photoproduction events. 
The data in the range 0.53 GeV/c$^2 < M_{\pi\pi} < 1.2$ GeV are fit to the sum of two Breit-Wigner distributions, for the $\rho$ and $\omega$, and a $\pi^+\pi^-$ continuum: 
\begin{equation}
 \frac{d\sigma}{dM_{\pi^{+}\pi^{-}}} \propto \left | A_{\rho}\frac{\sqrt{M_{\pi\pi}M_{\rho}\Gamma_{\rho}}}{M_{\pi\pi}^{2}-M_{\rho}^{2}+iM_{\rho}\Gamma_{\rho}}+B_{\pi\pi}+C_{\omega}e^{i\phi_{\omega}}\frac{\sqrt{M_{\pi\pi}M_{\omega}\Gamma_{\omega}}}{M_{\pi\pi}^{2}-M_{\omega}^{2}+iM_{\omega}\Gamma_{\omega}} \right |^{2} + f_{p}
\label{massFitFunction}
\end{equation}
where $\Gamma_\rho$ and $\Gamma_\omega$ are the mass-dependent widths for the two mesons:
\begin{equation}
\Gamma=\Gamma_V \frac{M_V}{M_{\pi\pi}} \bigg( \frac{M^2_{\pi\pi} - nm^2_\pi}{M^2_V - nm^2_\pi} \bigg)^{3/2}
\end{equation}
where $M_V$ and $\Gamma_V$ are the pole mass and width respectively, $m_\pi$ is the pion mass, and $n=2$ for the $\rho$ and $n=3$ for the $\omega$.  These widths account for the increase in phase space with increasing pair mass; the $\omega$ formula is based on the dominant $3\pi$ decay.  $A_\rho$, $B_{\pi\pi}$ and $C_\omega$ give the amplitudes for the $\rho$, direct $\pi\pi$ and $\omega$ respectively; we include the small $\omega\rightarrow\pi^+\pi^-$ branching ratio in $C_\rho$.
 $f_p$ is a quadratic polynomial which accounts for the remaining background.
 The fit has $\chi^2/DOF=314/297$.   

Although the $\omega$ component is small, $\rho$/$\omega$ interference intoduces a kink into the lineshape near $M_{\pi\pi}=M_\omega$.    The $\rho$ to direct $\pi\pi$ amplitude ratio $|B_{\pi\pi}/A_\rho|=0.79\pm 0.01 ({\rm stat.}) \pm 0.08 ({\rm syst.})$(GeV/c$^2$)$^{-1}$ is consistent with previous results from STAR \cite{Abelev:2007nb,Agakishiev:2011me}, HERA 
\cite{Aid:1996bs} and ALICE  \cite{Adam:2015gsa}.  The $\rho$ to $\omega$ amplitude ratio, $C_\omega/A_\rho = 0.049 \pm 0.0054 ({\rm stat.}) \pm0.0048  ({\rm syst.})$ is close to the
only previous measurement of $\rho/\omega$ interference in the $\pi^+\pi^-$ final state, $C_\omega/A_\rho=0.044\pm0.004$ \cite{Alvensleben:1971hz}.
This previous measurement by a MIT-DESY group was at a much lower beam energy, with 5-7 GeV photons striking a fixed target.  At that energy,  photon-meson fusion contributed significantly to the cross-section, so it is not obvious that one expects good agreement.  The fitted $\omega$ phase angle was non-zero, $1.73\pm 0.13 ({\rm stat.}) \pm  0.17 ({\rm syst.})$ radians, again in agreement with the DESY-MIT results; the presence of photon-meson fusion did not noticeably change the phase angle. 

The fit finds $M_\rho=0.7757\pm0.006$ GeV/c$^2$, $\Gamma_\rho=0.1475\pm0.0014$ GeV/c$^2$ and $M_\omega=0.7838\pm0.0009$  GeV/c$^2$, $\Gamma_\omega= 0.0163\pm0.0017$  GeV/c$^2$.  Except for $\Gamma_\omega$, which includes a contribution due to detector resolution, all are consistent with the standard values.

\begin{figure}[t]
\includegraphics[width=0.48\columnwidth]{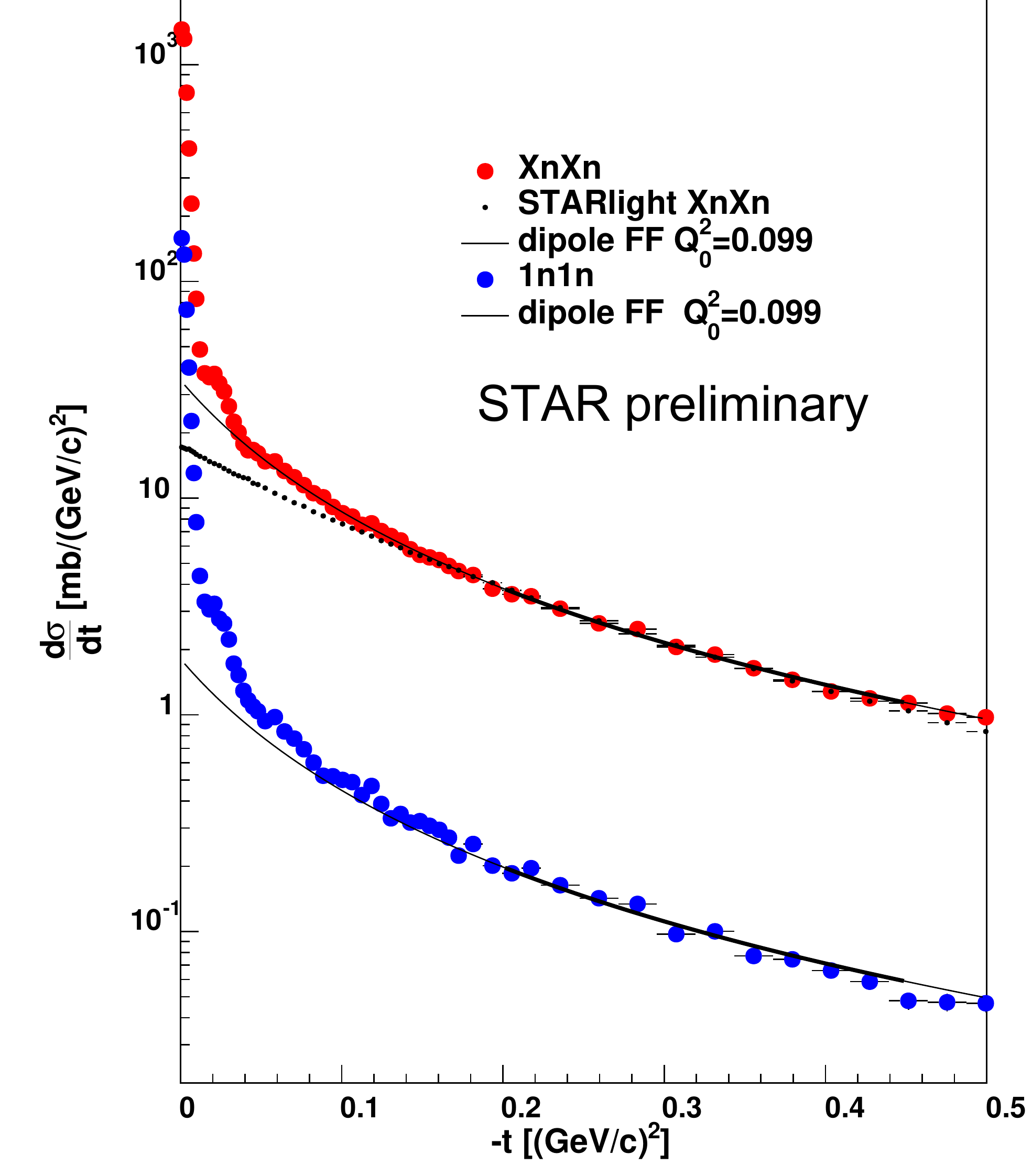}
\includegraphics[width=0.48\columnwidth]{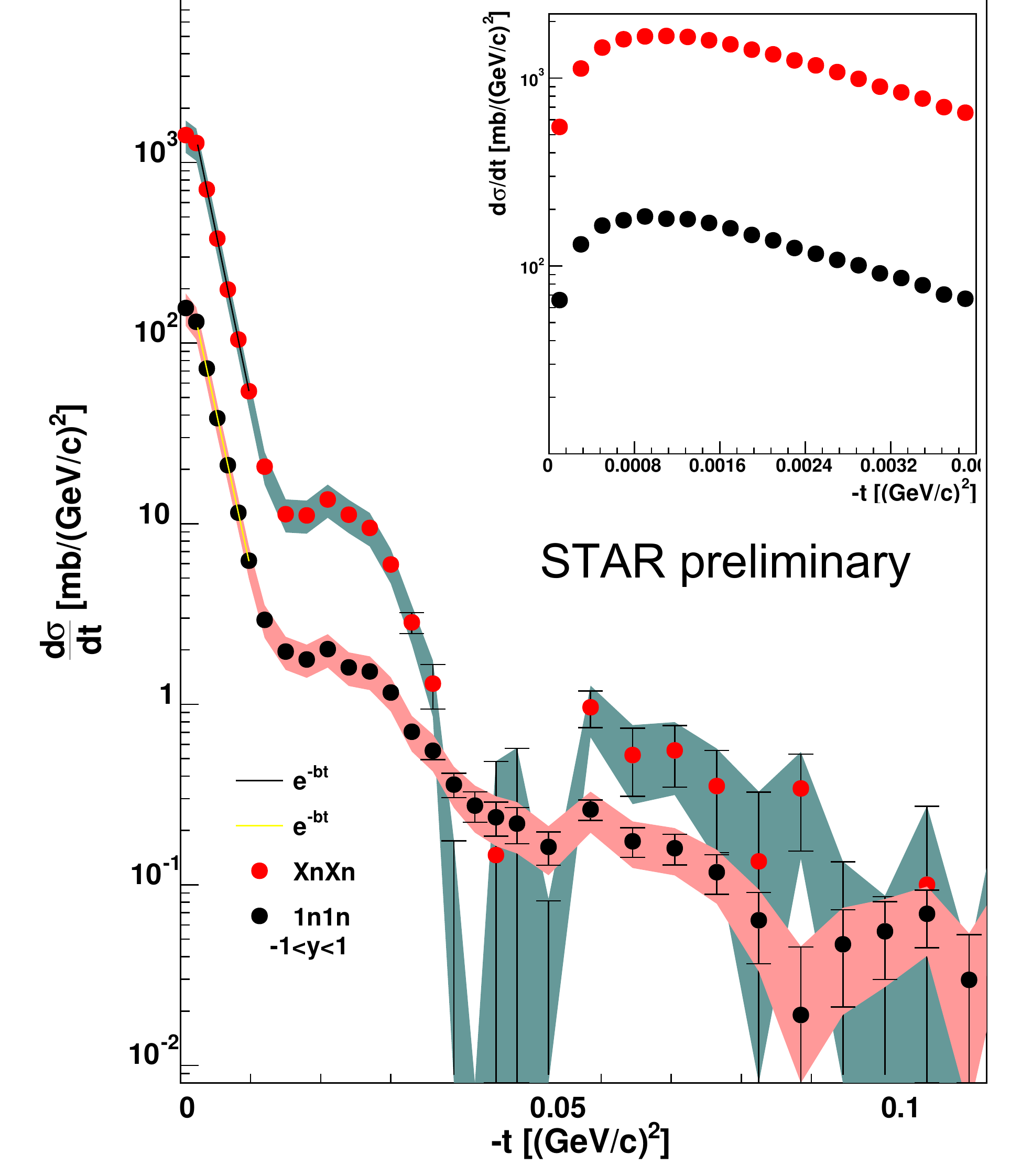}
\label{fig:dsdt}
\caption{(Left)  STAR preliminary $d\sigma/dt$ for exclusive $\rho^0$ photoproduction for events with a single neutron in each ZDC (1n1n) and any number of neutrons (XnXn) detected in each ZDC, along with fits of the incoherent contribution to a dipole form factor. (Right) the coherent contribution to $d\sigma/dt$ for 1n1n and XnXn interactions.  Two diffractive minima are visible for both selections. The error bars are statistical only.}
\end{figure}

Figure \ref{fig:dsdt} (left) shows $d\sigma/dt$ for two data samples, one with one neutron in each ZDC (`1n1n') and the other allowing any number of neutrons in each ZDC (XnXn). Traditionally, $d\sigma/dt$ could be broken up into coherent  (where the nuclei stays intact) and incoherent (where the nucleus breaks up) components.  The requirement for mutual Coulomb excitation in our trigger precludes this method of separation so we instead separate the two components spectrally.    We measure the incoherent contribution at large $-t$, by fitting
$d\sigma/dt$ in the range $0.2 {\rm GeV}^2 < -t < 0.45  {\rm GeV}^2$ to a dipole form factor.  This is then subtracted from the overall $d\sigma/dt$, leaving the coherent contribution, shown in in Fig. \ref{fig:dsdt} (right).  Two clear diffractive minima are visible; their positions are sensitive to the size of the target.  The dip for $|t|<0.001$ GeV$^2$, visible in the inset, is due to destructive interference between photoproduction at the two nuclei as $p_T\cdot \langle b\rangle\rightarrow 0$\ \cite{Abelev:2008ew}, where $\langle b\rangle$ is the median impact parameter.  One can apply a two-dimensional Fourier-Bessel transform to determine the two-dimensional (integrated in $z$) distribution of the interaction sites within the target nucleus.  

\section{A high mass final state}

\begin{figure}[t]
\center{\includegraphics[width=0.55\columnwidth]{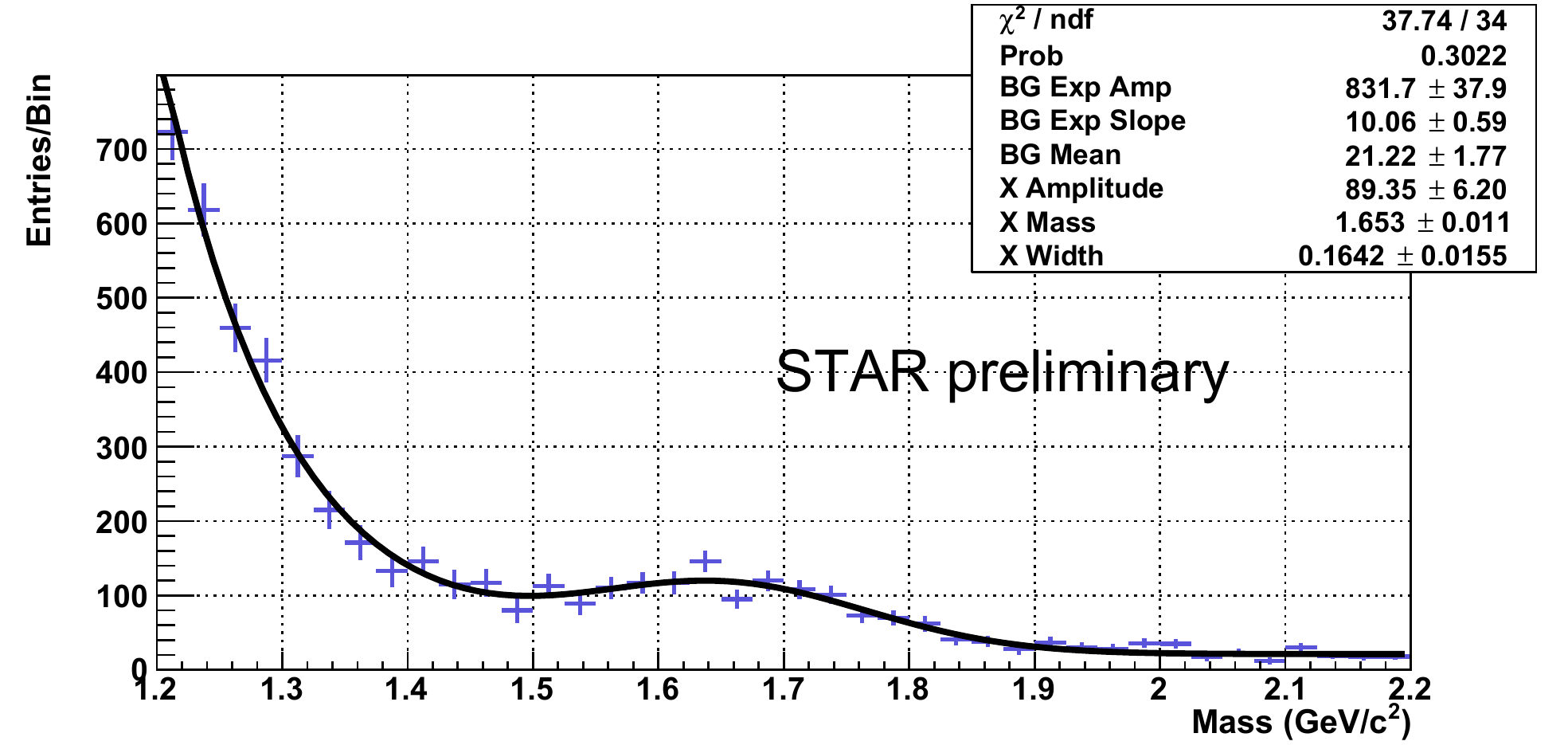}}
\label{fig:highmass}
\caption{
STAR preliminary dipion mass spectrum for pairs with $M_{\pi\pi}>1.2$ GeV, along with a fit to an exponential tail of the $\rho^0$, a constant background, and a Gaussian peak. Errors are statistical only.}
\end{figure}

STAR has also studied high mass $\pi^+\pi^-$ final states, using an expanded data set collected in 2010 and 2011; this dataset has about twice the statistics as the 2010 data alone.  Figure \ref{fig:highmass} shows the background subtracted (unlike-sign minus like-sign) high-mass region ($M_{\pi\pi}>1.2$ GeV).  The data are fit to the sum of an exponential tail of the $\rho^0$ and a Gaussian peak, plus a constant residual background.  The Gaussian peak is fit to a mass of $1653 \pm 10$ MeV, with a width of $164\pm 15$ MeV (statistical errors only), with $1034\pm 71$ events in the peak; a significance of  $15\sigma$.   The fit finds a $\chi^2/DOF$ of 37.7/34; a similar fit without the Gaussian had a $\chi^2/DOF$ of 235/37.This $\pi\pi$ mass spectrum peaks in a different place than a ZEUS study of exclusive electroproduction of dipions at HERA
\cite{Abramowicz:2011pk}.

The peak mass and width are inconsistent with the high-mass $4\pi$ final states observed by STAR \cite{Abelev:2009aa} and ALICE  \cite{Mayer}.  The resonance is too heavy to be the $\rho'(1450)$, and rather light for the $\rho'(1700)$; in addition, both of the $\rho'$ states are likely to have a very small branching ratio to $\pi\pi$.   The mass and width are in reasonable agreement with  the $\rho_3(1690)$, which has a branching ratio to  $\pi\pi$ of $23.6\%$.  The observed ratio of $\rho_3(1690)\!:\!\rho^0$ is in at least rough agreement with the cross-section for $\gamma p\rightarrow \rho_3(1690) p$ measured by the OMEGA photon collaboration \cite{Atkinson:1985nz}.  If the final state is spin 3, then it will not interfere with the other, spin-1 contributors to the $\pi\pi$ mass spectrum. 

\section{Conclusions}

STAR has presented a new study of dipion final states, with 100 times higher statistics than previous STAR measurements.  In addition to the $\rho$ and direct $\pi\pi$ photoproduction, we observe the $\omega$ and a high-mass resonance which is consistent with the $\rho_3(1690)$.

We have measured $d\sigma/dt$ for $\rho^0$ photoproduction in the range up to $t=0.5$ GeV$^2$, and separated the spectrum into coherent and incoherent components.   Two diffraction minima are clearly visible in the coherent contribution.  

This work was supported in part by the US Department of Energy under contract DE-AC03-76SF000098.

\end{document}